# Electron Wave Function Distribution Change in Mesoscopic Systems.

Erez Yahalomi
Department of condensed matter.
School of physics and Astronomy.
Raymond and Beverly Sacler Faculty of Exact Sciences.
Tel-Aviv University. Tel-Aviv 69978, Israel.

**We find a new phenomenon, a particle like an electron, which transfers kinetic energy to other subject undergoes a decrease in its wave packet size in space and an electron that gains kinetic energy experiences an enlargement of its wavepacket size. This effect occurs in some amount of degree almost in any physical system. The effect has significance on electron transport in semiconductors, quantum wires, future devices in nanotechnology and quantum optics.**

The effect is reveled in mesoscopic experiments of wave-particle duality, which for a century is the mysteriest question in quantum mechanics. There are several theories for explaining wave-particle duality such as, Bohr's interpretation [1] that the wave and particle aspects of quantum objects form a complementary pair or Bohm's theory [2], which, stated that a single quantum mechanical object consists of both an electron and an objectively real wave, which guide the electron. De Broglie's theory [3] argues a single electron consists of a physical wave solution in real space with a singularity, which gives rise to particle-like behavior. Development in experimental technology enables to study single electron effects. Such as experimental realizations of the two slit experiment [4-10] or the two path interferometer experiment [11-13], in these experiments the probability densities at a position x depends on the phase between the two wave function parts propagated through the two paths. Wave like behavior, or interference is possible when the different possible paths of a single electron can add coherently and thus interfere with each other, when a detector is activated the interference pattern is decreased . By analyzing these mesoscopic experiments we found a new main factor in addition to the decoherence factor [11-28] for the loss of interference in these experiments . In our new finding there is only a wave function or more accurately a wave packet that can change it size depending on the amount of the electron kinetic energy change. We find the electron wave packet decreasing its volume when transferring kinetic energy to external subject such as detector. When the wavepacket is decrease to sufficiently small size it exhibits in the detectors particle qualities: the electron wave packet become narrow enough to be detected as propagating mainly or only through one slit, resulting loss of interference pattern. For a detector that transfers energy to the electron wave function part, this electron wave function part gains kinetic energy causing it to change its main wave vector region and to increase its main distribution size. After this change the two wave function parts in the two different paths do not interfere constructively and loss of interference is obtained. We give here quantitive prediction for a one dimension case, of the magnitude of loss of interference in the two paths interferometer experiment. An electron that transfer kinetic energy due to interaction with another object, for example a detector based on energy absorption causing a signal detection on the detector when the detector absorbed energy from the electron. The detector transfers this kinetic energy to internal degrees of freedom or it can transport this energy to a screened region. For example an electron that interact with a detector based on a fluorescent screen, the electron scattered on the screen and transferred some of its kinetic energy to the atom on the screen, which excited by this energy and emitted light indicating detection of the electron. The electron loss some of its kinetic energy due to the inelastic scattering with the screen. This kinetic energy $\Delta E_k$ transferred to light energy and not to electron potential energy thus the total energy of the electron, $E_{tot}$ had reduced and this kinetic energy would never gain back to the electron since the photons produced from this kinetic energy propagated to the free space. The electron total final energy is changed to kinetic energy $E=E'-\Delta E_k$, where E' is the electron initial total energy. Schrodinger equation for this case is $(E'-\Delta E_k)\psi(x) = \frac{-h^2}{2m}\nabla^2\psi(x)$. The dynamical energy exchange with the detector is not appear in the equation, this make the equation simpler but general, since the kinetic energy loss is not limited to interaction with detector but can be any other physical phenomenon that cause a particle to loss its kinetic energy such as: electron moving in a crystal encountering a defect, electron influenced by external electric potential, electron collide by other particle in an accelerator.For a particle which gain kinetic from external source and has potential energy potential energy the equation is, $(E'+\Delta E_k - U)\psi(x) = \frac{-h^2}{2m}\nabla^2\psi(x)$. For a solution of the boundaries conditions of the transition between two states of initial kinetic energy and a decreased kinetic energy we use the mathematical tool of a potential step <u>that is constant in time as well</u>, it is identical to the solution of Schrodinger equation with constant potential $U=\Delta E_k$. Consider an electron wave packet encounters a potential step value of $U_0$, taking the electron kinetic as $E_k > U_0$. The energy change is shown in figure.1, figure.1a shows the potential energy of a propagating one-dimensional free electron wave packet encountering a potential step and transmitted through it. Fig .1b is the graph of the electron kinetic energy. The electron has a constant kinetic energy until it transmitted through the potential step where the electron's kinetic energy is reduced in an amount equals to the absolute amount of the potential energy increase $U_0$. A description by a potential barrier is not suitable since after a finite range a potential returns to zero potential and the kinetic energy is restored to the electron. While in a potential step the kinetic energy is not gain back to the electron presenting a constant kinetic energy loss and a one directional kinetic energy transfer from the electron. Except that in this general case the transmitted kinetic energy is not transformed to electron self potential energy.

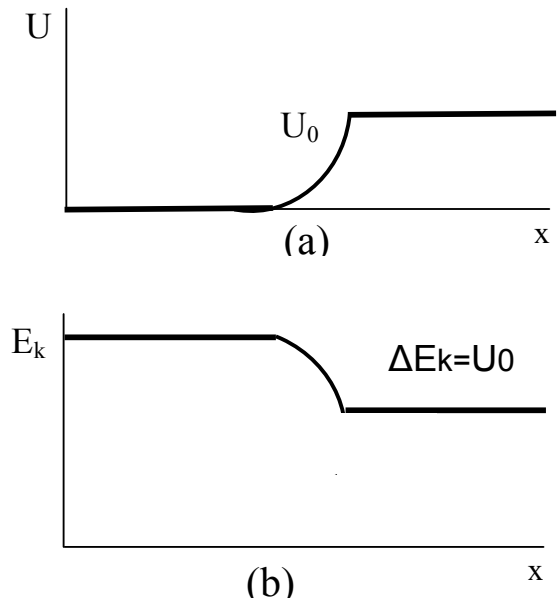

**Fig.1. (a) potential energy along the x axis; (b) The electron's kinetic energy loss along the potential energy rise.**

To demonstrate the effect in electron detection we look at the two slits experiment [4-10]. When the electron interacts with a detector that detect through which slit the electron propagate, the interference pattern decreases. Our finding is that kinetic energy is passed from the electron to the detector causing the electron wave function volume to decrease. The reduction in the electron wave function volume can happen even before the electron passes through one of the slits but has already passed energy to the detector for example by Coulomb repulsive potential, which slows the electron. This causes the electron wave function to be narrow enough to pass through only one slit, results in a loss of interfernce pattern and a particle behavior is detected by the detector since there is no wave function part going throught the second slit to interefere with the wave function part going through the first slit. We can illustrate this result by solving the two dimensional Schrodinger equation,

(1)
$$i\hbar \frac{\partial \psi(x,y,t)}{\partial t} = \frac{-\hbar^2}{2m}\frac{\partial^2 \psi(x,y,t)}{\partial x^2} + \frac{-\hbar^2}{2m}\frac{\partial^2 \psi(x,y,t)}{\partial y^2} + \Delta E_{1k}\psi(x,y,t) + \Delta E_{2k}\psi(x,y,t)$$

Where $\Delta E_{1k}, \Delta E_{2k}$ are the wave function kinetic energy loss in x,y axis respectively, we consider the two dimensions as independent and $Ek_{total} = E_{1k} + E_{2k}$. The free electron wave packet time dependent solution is $\psi_i$, where $\Delta E_{1k}=0$, $\Delta E_{2k}=0$ is,

$$\psi_i(x,y,t) = \frac{N_i}{2\pi} \int A(k_x) A(k_y) e^{i(k_x x + k_y y) - \frac{\hbar}{2m}(k_x^2 + k_y^2)t} dk_x dk_y$$

$k_x$ and $k_y$ are the wave vectors on x and y axis respectively. $N_t$ and $N_i$ are the normalization constants of $\psi_t$ and $\psi_i$ respectively. The probability weight coefficients are $A(k_x)$, $A(k_y)$ for x and y dimensions respectively, the wave packet is peaked in momentum space about the mean values $k_{0x,y}$. From x>a coordinate the electron kinetic energy loss value absorbed by the detector in x and y dimensions is $\Delta E_{1K}$ and $\Delta E_{2K}$ respectively, The electron wave packet solution $\psi_t$ after the kinetic energy loss is,

$$\psi_t(x,y,t) = \frac{N_t}{2\pi} \int A(k_x) A(k_y) e^{i(\sqrt{(k_x^2 - K_{0x}^2)}x - \frac{\hbar}{2m}(k_x^2 - K_{0x}^2)t + \sqrt{(k_y^2 - K_{0y}^2)}y - \frac{\hbar}{2m}(k_y^2 - K_{0y}^2)t)} dk_x dk_y$$

Where $K_{0x,oy} \equiv (2m\Delta E_{1k,2k}\hbar^{-2})^{0.5}$ for x and y axe respectively. We assume that when the kinetic energy transfer is complete there is no potential between the electron and the detector. If the initial kinetic energy of the wave vectors in the main region is larger than $\Delta E_{1K}$ and $\Delta E_{2K}$ the reflected part of the wave packet is negligible compared to the transmitted part Which proved by matching the boundaries condition of : a. wave function continuity. b. wave function derivative continuity, and assuming there is no incident wave function in the opposite direction from ∞ to (-)∞. For the x axe we obtain that the transmitted wave function or wave vector k is multiplied by the factor $\dfrac{2k_x}{k_x+\sqrt{(k_x^2-K_{0x}^2)}}$ , when $K_{0x}$ is smaller than $k_x$ this factor is close to unity. For the y axis we get $A'(k_y)=A(k_y)\dfrac{2k_y}{k_y+\sqrt{(k_y^2-K_{0y}^2)}}$ .

Another reason is most of the wave packet probability is centered in a narrow range around a certain wave vector, thus we can approximate that each transmitted wave vector have the same weight coefficient as before the kinetic energy transfer. Although A(k) does not change after the energy change, each of the A(k) weight coefficients relate after the energy change to a different wave vector. Instead of the wave vector k relates to the wave vector $(k^2-K_0^2)^{0.5}$. We expand around $k_{0x}$ and $k_{0y}$ by Taylor approximation $(k_{x,y}^2-K_{0x,y}^2)^{0.5} \approx q_{0x,y}+(k_{x,y}-k_{0x,y})k_{0x,y}\, q_{0x,y}^{-1}$ ,where $q_{0x,y} \equiv (k_{0x,y}^2-K_{0x,y}^2)^{0.5}$ . For the time evolution factor we do the expansion, $k_{x,y}^2-K_{0x,y}^2 = k_{0x,y}^2-K_{0x,y}^2+2*k_{0x,y}*(k_{x,y}-k_{0x,y})$. Substituting these gives,

(2)
$$|\psi_t(x,y,t)| \cong |N'\psi_i(\dfrac{k_{0x}}{q_{0x}}x-V_xt,\dfrac{k_{0y}}{q_{0y}}y-V_yt,0)|$$

where $N`=N_t/N_i$. From eq.2, we get the relation between the incident wave packet area $\Delta s_i = \Delta x_i \Delta y_i$ and the transmitted wave function area,

(3)
$$\Delta s_t = \dfrac{q_{0x}}{k_{0x}}\dfrac{q_{0y}}{k_{0y}}\Delta s_i$$

We find that the transmitted wave function area $\Delta s_t$, reduced due to the electron loss of kinetic energy. The particle wave function kinetic energy decrease loss is expressed by a change in the wave vector value as seen in eq.7. The width of a wave packet is determined by interference between the wave packet wave vectors. The main range of the wave vectors is $[k_1,k_2]$, when the particle losses kinetic energy as in various detection processes the wave vectors change. A kinetic energy decrease of $\Delta E$ will change the wave vector k to $(k^2-K_0^2)^{0.5}$ .The main wave vectors range changes to $[(k_1^2-K_0^2)^{0.5},(k_2^2-K_0^2)^{0.5}]$, which is broader than the initial range $[k_1,k_2]$. A broader range of wave vectors which substantially influence the interference inside the particle wave packet result in a decrease of the particle volume distribution.

Figure.2 shows an electron wave packet area calculated from eq.3 as a function of different final electron kinetic energy values, resulting from different amounts of kinetic energy transfer. The kinetic energy transferred is equals in x and y dimensions. The initial electron main wave function area is 2 nanometer$^2$ and the initial kinetic energy is 0.03 eV .

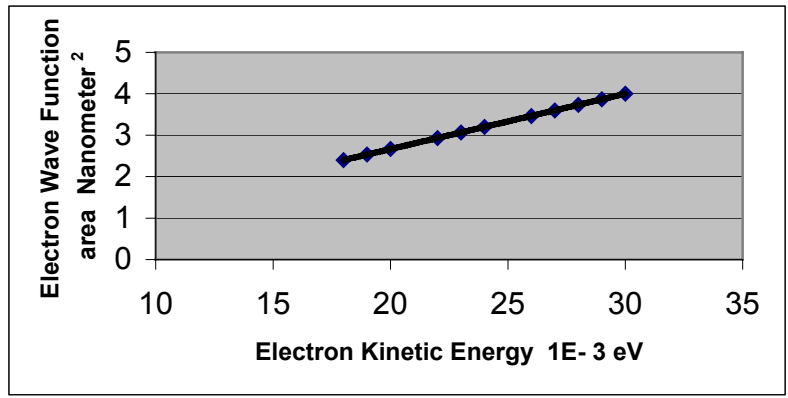

**Figure 2. electron wave function distribution area as a function of the electron kinetic energy.**

This size reduction can causes the electron wave function to become narrow and to propagate through only one slit, which explains the loss of interference pattern in the two slits experiment and the continues transition to isotropic pattern presenting electron that propagate through only one slit. In equations 2 and 3 when an electron loss kinetic energy the electron does not have to lose coherence. We find this is supported by an experiment [29] of multi-sodium atom coherent beam propagated through interference slits . When the coherent sodium atoms beam is scattered by photons, the photons changed the x component of the atoms momentums resulting a loss of the interference pattern. When detecting only a narrow part of the scattered beam, which relates to atoms that went through the same momentum change due to the photon scattering, most of the interference pattern restored. Hence two electrons that were coherent before kinetic energy change are still coherent after they experienced a similar kinetic energy change. In order that the multi-atom beam would be coherent each single atom in the beam should be coherent. This lead us to the same result we obtained from the equations that a single electron remains coherent after kinetic energy change.

From equation 2 it is obtained that the magnitude of the electron wave function size change depends on the electron initial kinetic energy. This quality can be used in nanotechnology devices: on a strong noisy environment the initial electronic wave function would by higher making the nano wave function less sensitive to kinetic energy changes from noise, on a less noisy environment the initial kinetic energy of the electronic wave function can be smaller, for achieving significant wave function size change by smaller change in the electron kinetic energy.

For explanation of the two slits experiment with a detector type that emits energy signal and by the change in the emitting signal detect the electron location, we look at the two paths interferometer experiment [11-13]. In this experiment an electron wave function propagate through two separate paths , the paths have joint beginning and ending. This experiment is different from the two slits experiment in a manner that when the detector is activated on one path there is almost no connection between the two wave function parts in the two different paths. A quantum point contact detector [30], is near one of the two paths we deduce this detector transmits kinetic energy to the measured electron through the coulomb interaction between the electron in the path and the electron current in the detector, we shall show that the energy gain in the observed electron caused loss of interference.

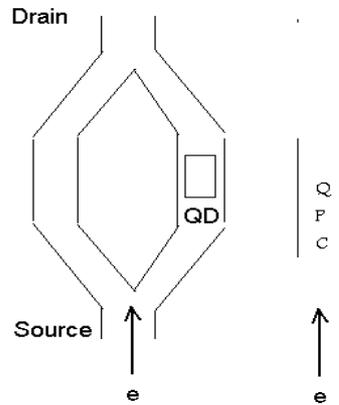

**Fig.3. Two path interferometer.** QD is quantum dot, QPC is the detector.

We write the single electron wave function as a product of the wave function confined at the boundaries potential in the transverse y direction, and a free wave function along the propagation x direction. The wave function is split in the two interferometer paths:

$$\psi_1(x,y,t) = A_1\phi(y,t)\frac{1}{2\pi}\int A(k_x)e^{i(k_x x - \frac{\hbar}{2m}k_x^2 t)}dk_x$$

$A_1$, $A_2$ are the probability amplitudes of $\psi_1$ and $\psi_2$, x,y internal axis inside each of the paths. We looked at the interference along the x direction between the two wave function parts after exiting the two paths, the kinetic

$$\psi_2(x,y,t) = A_2 \phi(y,t) \frac{1}{2\pi} \int A(k_x) e^{i(k_x x - \frac{\hbar}{2m}k_x^2 t)} dk_x$$

energy has gained in the electron wavefunction part that interacted with the detector. This is deduced by analyzing the experimental results [13], indicating a direct relation between increase in the detector conductance and increase in the QD conductance and the observed electron inside it, which result in decrease in the interference visibility of the measured electron. The decrease of the visibility in the experiment is continues [13], this is different from earlier theories, which a particle detection assumed to cause a discrete change from wave interference pattern to particle interference pattern. In a similar experimental setup [28], electromagnetic wide band shot noise in frequency range of 8-18 GHz, associated with DC current through the QPC was measured, this electromagnetic energy interact with the observed electron in the QD, which indicate the electron gain energy. The end of the interaction region is at x>a we neglect the coulomb potential after coordinate region: a, where the detector's electrons charge is screened. We consider the electron wave vector main region is around $k_0$ and $k_0 > K_0$,

$$\frac{2k_x}{k_x + \sqrt{(k_x^2 + K_{0x}^2)}}$$ is the change in the weight function obtained from

boundaries conditions for $K_{0x} < k_x$ this term is approximately 1.

To obtain the electron wave packet part expanded due to kinetic energy gain we calculate by Taylor series, $(k_x^2 + K_{0x}^2)^{0.5} \approx p_0 + (k_x - k_{0x})*k_{0x}*P_0^{-1}$, where $p_0 = (k^2_{0x} + K^2_{0x})^{0.5}$. We also expand the time evolution part and we get,

$$\psi_1(x,y,t) = \frac{1}{2\pi} A'_1 \phi'(y) e^{i(p_0 x - \frac{\hbar}{2m}(k_{0x}^2 + K_{0x}^2)t)} \int \frac{2k_x}{k_x + \sqrt{(k_x^2 + K_{0x}^2)}} A(k_x) e^{i(k_x - k_{0x})(\frac{k_{0x}}{p_{0x}}x - Vt))} dk_x$$

$A'_1$ is normalized amplitude factor. We obtain relation between the initial wave function width $\Delta x_i$ and the wave function width after gaining kinetic energy $\Delta x_t$: **$\Delta x_t = p_0 * k_{0x}^{-1} * \Delta x_i$.** We found the increase of the electron's kinetic energy reduce the main wave vectors region and increase the electron wave packet width. After the electron wave packet increase due to receiving kinetic energy from the detector's electrons current, there is still interference between the two paths where the wave function in the path with no detector is unchanged. The interference between the two wave packet parts consisting of two different main wave vector ranges and different longitude widths cause decrease in the interference pattern. The interference equation for the x axis component of the interference is,

(4)
$$P_x = |\psi_{1x}|^2 + |\psi_{2x}|^2 + 2\operatorname{Re} A'_{1x} A^*_{2x} \int\int A(k_{1x}) A^*(k_{2x}) \frac{2k_{1x}}{k_{1x} + \sqrt{k_{1x}^2 + K_{0x}^2}}$$
$$\cos((\sqrt{k_{1x}^2 + K_{0x}^2}) - k_{2x})x - \frac{\hbar(k_{1x}^2 - K_{0x}^2 - k_{2x}^2)}{2m} t) dk_{1x} dk_{2x}$$

$P_x$ is the x component of the probability density, $\psi_{1x,2x}$ are the x component of the electron wave function parts $\psi_{1,2}$ respectively $A_{1x,2x}$ are the x component of the probability amplitudes of $A_{1,2}$. $k_{1x,2x}$ are the x component of $\psi_{1,2}$ wave vectors respectively, $A(k_{1x,2x})$ are the propability weight cooeficients. The third term in eq.4 is the interference term between the wave vectors of the two wave functions, cause reduction of the inference pattern visibility. The interfernce in y axis can be obtained in a similar way with consideration of the bounderies condition in the path . The total propability is a product of $P_x$ and $P_y$.

Althought the uncertainty relation can give an approximation for the wavepacket length when the momentum distribution of the wave packet is given. The relation presented in this paper is between kinetic energy difference and location distribution and not between two uncertainies. In the two slits experiment the explanations by the uncertainty relation derived from the quantum character of the measurement device, like the device sensitivity to momentum, which cause uncertainty in the device position or from the slit width which due to the width of the slit leads to a back action, we calculate the backaction according to the data of experiment [28] and found that back action by uncertainty relation is seven orders of magnitude smaller than electron energy gain in the detection process ,so it can not explain the loss of interfernce . In the two slit experiment we even found that reduction of the elecron wave function area due to detection can be before the electron enter the slits, by a long distance transmission of energy from the electron to the detector, for example a Couloumb potential which slows electron.

We conclude, in mesoscopic experiment where electron transmit energy the electron wave function described as a wave packet reduced its main distribution size in space . This size reduction cause the obderved electron wave packet in the two slits experiment to propagate mainly or only through one slit result in loss of interference pattern. The magnitude of the electron wave packet size change is correlated to the amount of kinetic energy loss. When the electron gains kinetic energy the electron wave function space distribution increases.The elctron does not have to change it's wave behavior it changes its size in space. In future papers we intend to explore the effect in many electrons system. The effect estimated to has significant influence on conductivity of semiconductors and quantum wires. This effect can give solution for cases where coherence electron propogation is tried to be applied and encounterd difficulties of rapid coherence loss, since this phenomnon does not depends essentialy in coherence and as we find the electrons do not have to loss coherene when they change there kinetic energy. This effect in mesoscopic systems may lead to new devices in nano technology using this effect for incresing and decreasing the global electronic wave function volume of nano devices. From primary caloclations we obtain that a similiar effect acours also in a photon in future research we intend to probe the effect in quantum optics.

**Aknowledgment**
The arthur would like to thank L.P Horwitz , G. Kventsel , M. Moshe , A. Ron and A. Stern for useful suggestions. I like to thank Y.Aharonov for presenting me the very interesting subject of wave particle duality in his courses. The paper is based on a research I started in 1997 during the M.Sc. studies.